\documentclass[aps,pre,showpacs,twocolumn]{revtex4}
\usepackage{eurosym}
\usepackage{amssymb}
\usepackage{graphicx}
\usepackage{colordvi}
\usepackage{color}
\usepackage{dcolumn,amsmath,amsthm,amscd,amsfonts,amssymb,epsfig,graphics,graphicx,eucal}
\usepackage{epstopdf}

\begin{document}

\title{Nonlinearity-induced localization in a periodically-driven
semi-discrete system}

\author{R. Driben$^{1}$, V. V. Konotop$^{2}$, B. A. Malomed$^{3,4}$, T. Meier$^{1}$, and A. V. Yulin$^{4}$}
\affiliation{
$^{1}$Department of Physics and CeOPP, University of Paderborn, Warburger Str. 100, D-33098 Paderborn, Germany\\
 $^{2}$ Centro de F\'{i}sica Te\' orica e Computacional  and Departamento de F\'{i}sica, Faculdade de Ci\^encias, Universidade
de Lisboa, Campo Grande, Ed. C8, Lisboa 1749-016, Portugal\\
 $^{3}$ Department of Physical Electronics, School of Electrical Engineering,
Faculty of Engineering, and Center for Light-Matter Interaction, Tel Aviv
University, P.O.B. 39040, Ramat Aviv, Tel Aviv, Israel\\
$^{4}$ITMO University, 49 Kronverskii Ave., St. Petersburg 197101, Russian Federation}
%\author{Boris A. Malomed$^{7,8}$}
%\affiliation{Department of Physical Electronics, School of Electrical Engineering, $^{7}$%
%Faculty of Engineering, and Center for Light-Matter Interaction, Tel Aviv
%University, P.O.B. 39040, Ramat Aviv, Tel Aviv, Israel}
%\affiliation{$^{8}$ITMO University, St. Petersburg 197101, Russia}
\date{\today }

\begin{abstract}
We demonstrate that nonlinearity plays a constructive role in supporting the robustness
of dynamical localization in a model which is discrete, in one dimension and continuous in the
orthogonal one. In the linear regime, time-periodic modulation of the gradient strength along the discrete axis
leads to the usual rapid spread of an initially confined wave
packet. Addition of the cubic nonlinearity makes the dynamics drastically
different, inducing robust localization of moving wave packets. Similar
nonlinearity-induced effects are also produced by combinations of static and
oscillating linear potentials. The predicted nonlinearity-induced dynamical
localization can be realized in photonic lattices and Bose-Einstein
condensates.
\end{abstract}

\maketitle

%\title{Nonlinear dynamics of spatio-temporal wave packets experiencing Bloch oscillations in arrays of nonlinear fibers with high order dispersion. Emission of resonant radiation.}

%\author{R. Driben}
%\affiliation{Department of Physics and CeOPP, University of Paderborn, Warburger Str. 100, D-33098 Paderborn, Germany}
%
%\author{T. Meier}
%\affiliation{Department of Physics and CeOPP, University of Paderborn, Warburger Str. 100, D-33098 Paderborn, Germany}

\input{epsf.tex} \epsfverbosetrue

\section{Introduction}

The possibility of Bloch oscillations (BOs)~\cite{Zener}, i.e., the
occurrence of a temporally-oscillating (ac) electrical current originating
from spatial oscillations of the electron charge density in a crystal biased
by a static uniform (dc) electric field, in the absence of scattering
effects, was predicted by Bloch and Zener almost 90 years ago. Being
initially far from the reach of feasible experimental realizations, this
prediction caused debates regarding the actual existence of BO, which lasted
for several decades. The proof securing that BO should be physically
realizable, as predicted by effective Hamiltonians that include a finite
number of bands, was theoretically provided in the early 1990s when rigorous
upper limits for the interband tunneling rates had been established, see,
e.g., Ref. \cite{nenciu}. Also in the 1990s, BOs had been first observed
experimentally in the temporal domain in electrically-biased semiconductor
superlattices using optical interband excitation by femtosecond laser pulses~%
\cite{jochen}. A few years later, BOs have been realized with ultracold
atoms in optical lattices~\cite{atoms} and successfully emulated in optics,
using arrayed waveguides~\cite{Peschel,Silberberg}. These results prove that
BOs are a general physical effect relevant for a large class of systems.

In Refs. \cite{Peschel,Silberberg} it was shown that, if the phase velocity
of the waves varies linearly with discrete coordinate $n$ of waveguides in
the array, the position of the light beam is an oscillating function of
propagation distance $z$, which is the optical counterpart of the electronic
BO dynamics. In the latter context, the influence of the optical Kerr
nonlinearity was considered too. However, in models of arrayed waveguides,
with only discrete diffraction, the nonlinearity was shown to produce a
destructive effect on the BO dynamics. On the other hand, it was shown in a
recent work \cite{Driben_BlochOscill} that adding another dimension with
continuous diffraction may result in a constructive effect of the
nonlinearity, \textit{viz}., localization of the wave packet in space and
the emergence of a quasi-solitonic regime of propagation. Thus, one may
expect the existence of a new species of robust nonlinear hybrid wave
packets, combining features of solitons and Bloch-oscillating waves. In
photonic systems, the robustness of the hybrid wave packets in the presence
of the anomalous group-velocity dispersion may lead to prediction of
resonant radiation with nontrivial properties, similar to how it was
predicted in the spatial domain under the action of diffraction \cite{Radiation}.
This methodology is conceptually different from several proposed ways to guide nonlinear waves experiencing BOs in nonlinear regime that
include variation of nonlinear interactions strength along the evolution \cite{soliton1,soliton2}.

A phenomenon somewhat related to BOs, which may be induced by time-modulated
gradient potentials, is dynamical localization (DL). It was predicted in
Ref.~\cite{Dunlap} within the framework of the tight-binding approximation
for electrons in solids. Whereas usually an initially localized wave packet
will delocalize when driven by an oscillating bias, the DL implies that the
wave packet remains localized. This is the case for a single-band
tight-binding system, when the ratio of the amplitude of the bias and its
modulation frequency takes special values, namely, roots of Bessel function $%
J_{0}$. Thus, the conductance, provided for by the delocalization of the
wave packets, vanishes for such a special choice of the ac bias. Later, the
DL was studied theoretically for ultracold atoms trapped in optical lattices
\cite{Holthaus}, where it may be used for the coherent control of atoms and
for the realization of the superfluid Mott-insulator phase transition \cite%
{Holthaus2}. The same effect was also predicted to significantly alter the
optical absorption in semiconductor superlattices and the effective
dimensionality of excitons \cite{MeierDL}. DL has been experimentally
observed in atomic systems (see, e.g., Ref. \cite{raizen}), in transport
properties of semiconductor superlattices (see Ref. \cite{ucsb}), and it
also occurs in photonic settings \cite{Longhi}, where periodic
corrugation of waveguides makes it possible to optically emulate the
time-periodic linear force acting on a quantum particle.

 The main aim of the present work is to demonstrate that the DL persists in the nonlinear propagation
regime for hybrid soliton-BO wave packets in systems containing an
additional continuous dimension, in addition to the discrete one.
Additionally, we report results for the simultaneous presence of static and
oscillating gradient potentials. In the latter case, the electronic
transport is typically supported by multi-photon-assisted tunneling \cite%
{Ignatov,69,71,Meier}. As predicted in Ref.~\cite{Ignatov} for special
relations between the amplitudes and modulation frequency of the fields, one
can selectively suppress particular single- or multiple-photon transitions,
thus designing specific transport properties, as shown in Ref. \cite{Meier}
for semiconductor superlattices. We demonstrate that, in the semi-discrete
system considered here, these effects may persist in the presence of
significant nonlinearity.

The rest of the paper is organized as follows. In Sect.~\ref{model} we
introduce the model and present some considerations for it. Numerical and
analytical results are reported and discussed in Sect.~\ref{results}. The
paper by Sect.~\ref{concl}.

\section{The model and analytical results for the linear case}

\label{model}

We start with the semi-discrete model, introduced in Ref. \cite%
{Driben_BlochOscill} as the system of linearly-coupled Gross-Pitaevskii
equations (GPEs) \cite{Pit} for the Bose-Einstein condensate (BEC) loaded
into an array of waveguides/wires, written here in the normalized form:
\begin{gather}
i\frac{\partial u_{n}}{\partial t}+\frac{1}{2}\frac{\partial ^{2}u_{n}}{%
\partial x^{2}}+\kappa (u_{n-1}+u_{n+1}-2u_{n})  \notag \\
+\gamma (t)nu_{n}+g|u_{n}|^{2}u_{n}=0.  \label{GPE}
\end{gather}%
Here $u_{n}(x,t)$ is the mean-field wave function in the $n$-th waveguide, $%
\kappa $ is the array's coupling constant, $\gamma (t)$ is the strength of
the time-dependent gradient potential acting in the discrete direction
gradient, and the nonlinearity coefficient $g>0$ accounts for the intrinsic
self-attraction of the BEC.
For further numerical simulations we will fix values of $\kappa=2$ and $g=1$ for all the nonlinear simulations, while for the linear we will take $g=0$.
In terms of optics, Eq.~(\ref{GPE}), with $t$
and $x$ replaced, respectively, by the propagation distance, $z$, and
reduced time, $\tau $, is the system of coupled nonlinear Schr\"{o}dinger
equations (NLSEs) modeling the light propagation in an array of coupled
optical fibers in the presence of a linear gradient of the waveguide's
effective index \cite{Aceves}, whose magnitude may be modulated along $z$.
In the latter case, $u_{n}$ are scaled envelope amplitudes of the
electromagnetic waves in the fibers, the group-velocity dispersion is
anomalous, and the cubic terms with $g>0$ represents the Kerr nonlinearity.
Alternatively, the same system (\ref{GPE}), with $x$ being the transverse
coordinate, models the spatial-domain light propagation in a stack of
parallel planar waveguides, the second derivatives representing the paraxial
diffraction in the waveguides \cite{stack,Roy}.

Next, we consider the transport mechanism in the linear systems, with $g=0$
in Eq. (\ref{GPE}), keeping solely the discrete direction in it. For the
harmonic format of the time modulation of the gradient potential, with
\begin{equation}
\gamma (t)=\gamma _{0}\cos (\omega t),  \label{eq:modulation}
\end{equation}%
the DL is realized if condition
\begin{equation}
J_{0}\left( \gamma _{0}/\omega \right) =0,  \label{eq:zeroBessel}
\end{equation}%
with Bessel function $J_{0}$, is satisfied, as was shown in Ref. \cite%
{Dunlap,Holthaus}. The combined ac-dc-driven regime was analyzed in Ref.
\cite{Ignatov}, and realization of this effect was elaborated for the
photoexcited electronic transport in semiconductor superlattices in Ref.
\cite{Meier}. Based on those studies, one can predict that, for gradients of
the form
\begin{equation*}
\gamma (t)=\gamma _{\mathrm{dc}}+\gamma _{0}\cos (\omega t),
\end{equation*}%
the dynamical transport is suppressed if the following two conditions are
met simultaneously, for integer $m$:
\begin{equation}
\gamma _{\mathrm{dc}}=m\omega ,~J_{m}\left( \gamma _{0}/\omega \right) =0,
\label{m}
\end{equation}%
since in this case the $m$-photon-assisted tunneling vanishes for
tight-binding systems.

For an initial pulse that is smooth with respect to $n$, one can approximate
the finite difference in Eq. (\ref{GPE})\ by the continuum limit \cite%
{Kevrek}. The resulting two-dimensional (2D) linear continuum equation
admits at exact solution,
\begin{eqnarray}
u_{n} &=&\exp \left[ 2i\kappa \left( t+\int_{0}^{t}\cos (\Gamma (\tau
))d\tau \right) -in\Gamma (t)\right] \times   \notag \\
&&\times \frac{A\left( n+n_{0}(t)\right) \exp \left[ \frac{i}{2}A\left(
n+n_{0}(t)\right) t\right] }{\cosh \left[ A\left( n+n_{0}(t)\right) x\right]
},  \label{approxim_BO}
\end{eqnarray}%
where the location of the center of the pulse is given by
\begin{equation*}
n_{0}(t)=2\kappa \int_{0}^{t}\sin \left( \Gamma (\tau )\right) d\tau \text{,}
\end{equation*}%
with
  $
 \Gamma(t)=\int_0^t\gamma(\tau)d\tau,
 $ and $A(n)$ describes the pulse envelope in the input which gives rise to
solution (\ref{approxim_BO}):
\begin{equation}
u_{n}^{(0)}(x)=\frac{A\left( n\right) }{\cosh \left[ A\left( n\right) x%
\right] }.  \label{input}
\end{equation}%

As we demonstrate below, the semi-discrete model gives rise to a remarkably
robust DL\ regime in the nonlinear regime. In our numerical simulations the
same input (\ref{input}), which generates solution (\ref{approxim_BO}) in
the continuum limit, was taken, but in the semi-discrete format, with the
Gaussian envelope:%
\begin{equation}
A(n)=a_{0}\exp \left( -n^{2}/w^{2}\right) .  \label{Gauss}
\end{equation}%
Thus, we will consider the evolution of the input localized as $\mathrm{sech}
$ in the continuous coordinate, and as the Gaussian of width $w$ along the
discrete coordinate.

\begin{figure}[tbp]
\includegraphics[width=0.5\textwidth]{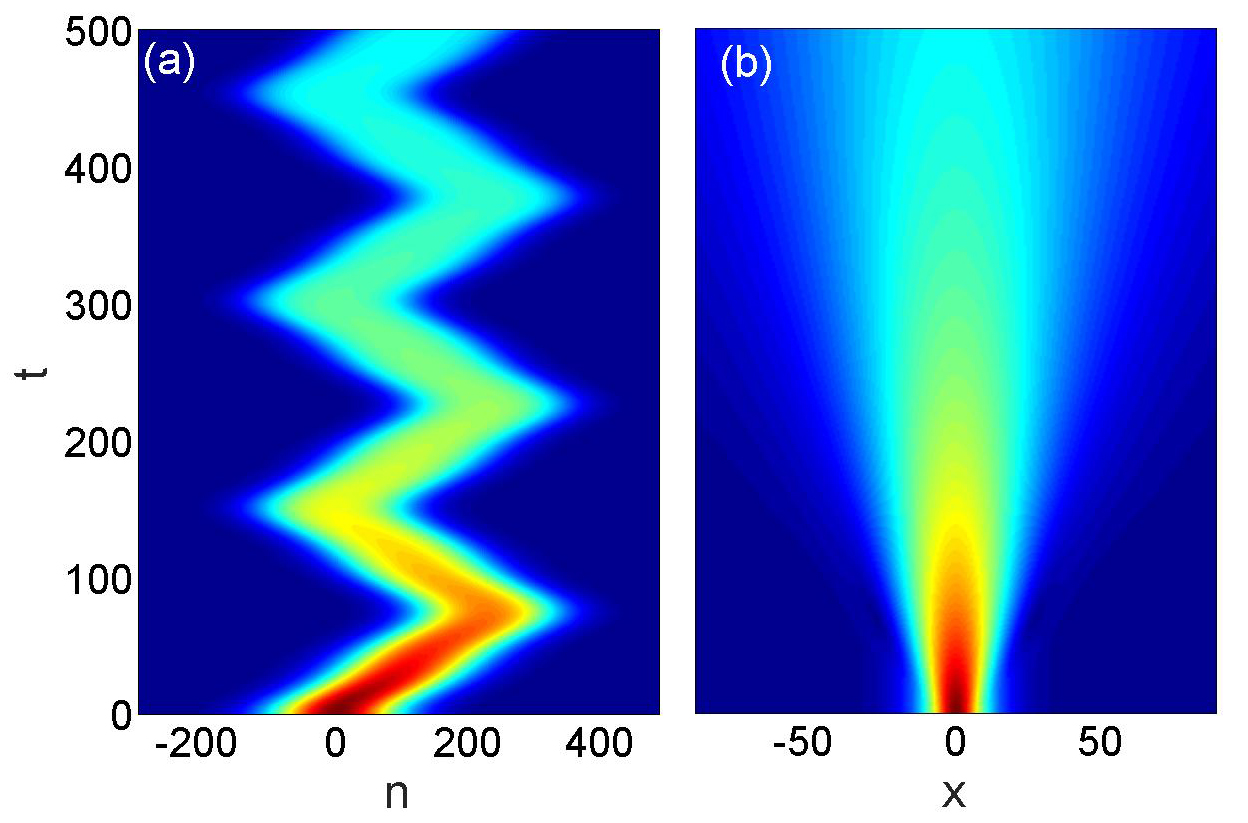}
\caption{(Color online) Evolution of the wave packet in the linear regime,
i.e., with $g=0$ in Eq.~(1), in the $\left( n,t\right) $ (a) and $\left(
x,t\right) $ (b) planes. The gradient strength is $\protect\gamma _{0}=0.1$
and the modulation frequency $\protect\omega =0.0416$, i.e., the ratio $%
\protect\gamma _{0}/\protect\omega \approx 2.4$ is very close to the first
root of the zeroth-order Bessel function $J_{0}$. Other system parameters: $\kappa=2$ and $g=0$. The initial condition is
given by Eqs.~(\protect\ref{input}) and (\protect\ref{Gauss}) with $%
a_{0}=0.15$ and $w=100$ (i.e., the initial Gaussian is quite broad).}
\end{figure}

\section{Results}

\label{results}

First we examine the evolution of the wave packet in the linear regime by
setting $g=0$ in Eq.~(\ref{GPE}). Figure~1(a) clearly shows that the wave
packet performs regular oscillations in $n$-direction with period $2\pi
/\omega $ of the modulation of the gradient potential. As we chose the
values of the gradient strength $\gamma _{0}=0.1$ and frequency $\omega
_{0}=0.0416$, which correspond to the first root of $J_{0}$ in Eq. (\ref%
{eq:zeroBessel}), the wave packet remains localized in the $n$-domain in the
course of the evolution, which is typical for the DL; however, the
localization degree is weakening with the increase of $t$. Concomitantly,
the diffraction along the continuous $x$-axis leads to rapid spreading of
the wave packet, as seen in Fig. 1(b). Thus, in the linear regime the wave
packet suffers delocalization in the course of its evolution, even when the
DL\ condition (\ref{eq:zeroBessel}) holds.

\begin{figure}[tbp]
\includegraphics[width=0.5%
\textwidth]{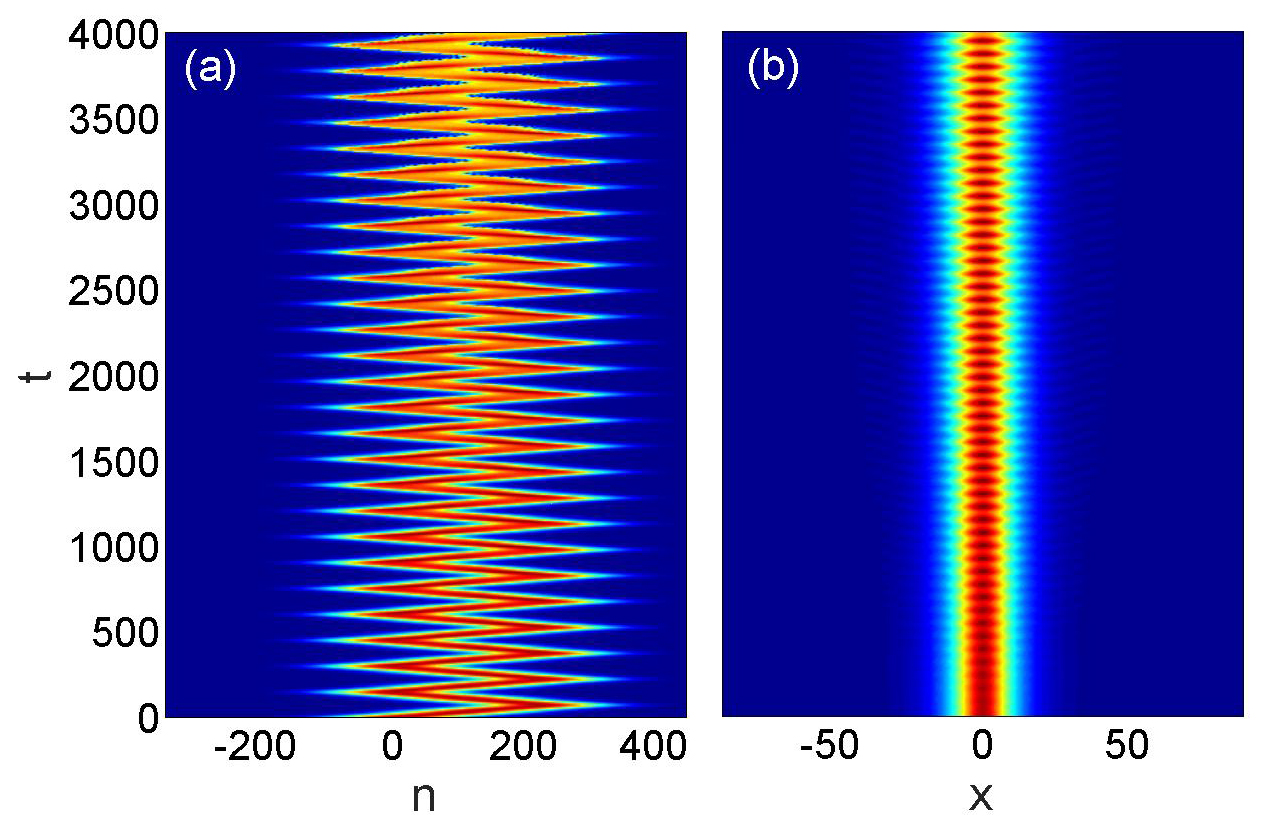} %
\caption{(Color online) Long-time evolution of the wave packet in a
nonlinear system in the $n$-$t$ and in the $x$-$t$ planes, respectively. (a)
and (b) Dynamical localization of the wave packet driven with a strength and
frequency corresponding to the first root of the Bessel function $J_{0}$.
The parameters are chosen as $\protect\gamma =0.1$ and and $\protect\omega %
=0.0416$. The two other parameters are taken: $\kappa=2$ and $g=1$ as for the rest nextcoming numerical simulations.The initial condition is given by Eqs.~(\protect\ref{input}) and
(\ref{Gauss}) with $a_{0}=0.15$ and $w=100$.\newline}
\end{figure}

Next we launch the input into the full nonlinear system, corresponding to
Eq.~(\ref{GPE}) with $g=1$. The other parameters are same as in Fig.~1,
i.e., the ratio of the gradient strength and frequency corresponds to the
first root of $J_{0}$ in Eq. (\ref{eq:zeroBessel}). Figures~2(a) and (b)
display the nonlinear evolution in the $\left( n,t\right) $ and $\left(
x,t\right) $ planes, respectively. Now, instead of the gradual decay of the
linear wave packet displayed in Fig.~1, long-lived robust localization of
the wave packet is observed, in both the $n$- and $x$-directions. We thus
conclude that the nonlinear propagation creates self-trapped robust 2D
dynamical semi-discrete soliton like structures, impossible in the linearized system.

Dynamical localization is also found when frequency $\omega $ is reduced so
that the ratio of $\gamma $ and $\omega $ corresponds to the second root of $%
J_{0}$ in Eq. (\ref{eq:zeroBessel}), as shown in Figs.~3(a) and (b).
However, in a still stronger nonlinear regime, corresponding to the
amplitude $a_{0}$ which is twice as large, the robustness of the established
wave packets drops significantly, as we reach the quasi-collapse-driven
dynamical regime \cite{Kevrek}, with the wave packet splittings, similar
to what was reported in Ref. \cite{Driben_BlochOscill}. Figures~3(c) and (d)
illustrates this situation. Thus, similar to the case of the BO, cf. Ref.
\cite{Driben_BlochOscill}, the well-pronounced DL regime occurs at optimal
strengths of nonlinearity, which can be identified by systematic simulations
with varying $a_{0}$ and/or $g$.

\begin{figure}[tbp]
\includegraphics[width=0.5%
\textwidth]{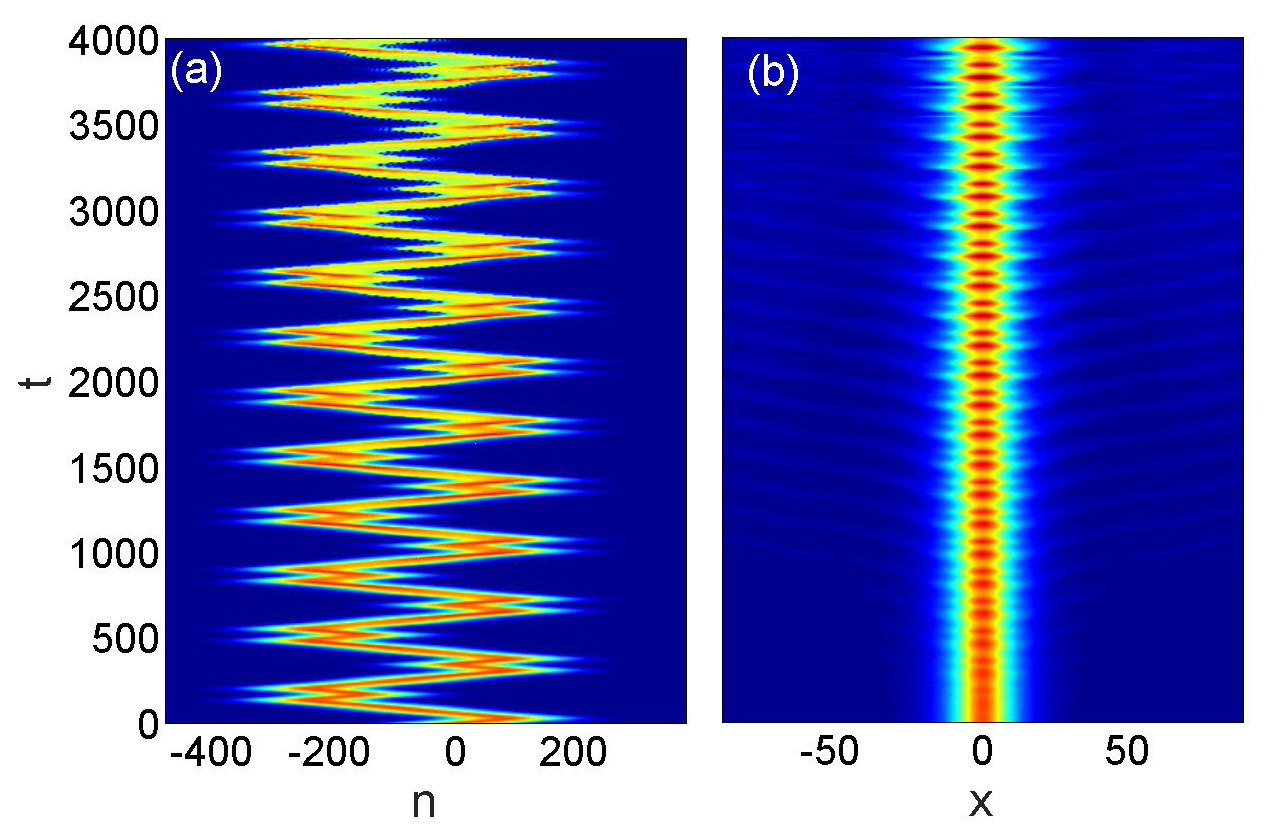} %
\includegraphics[width=0.5\textwidth]{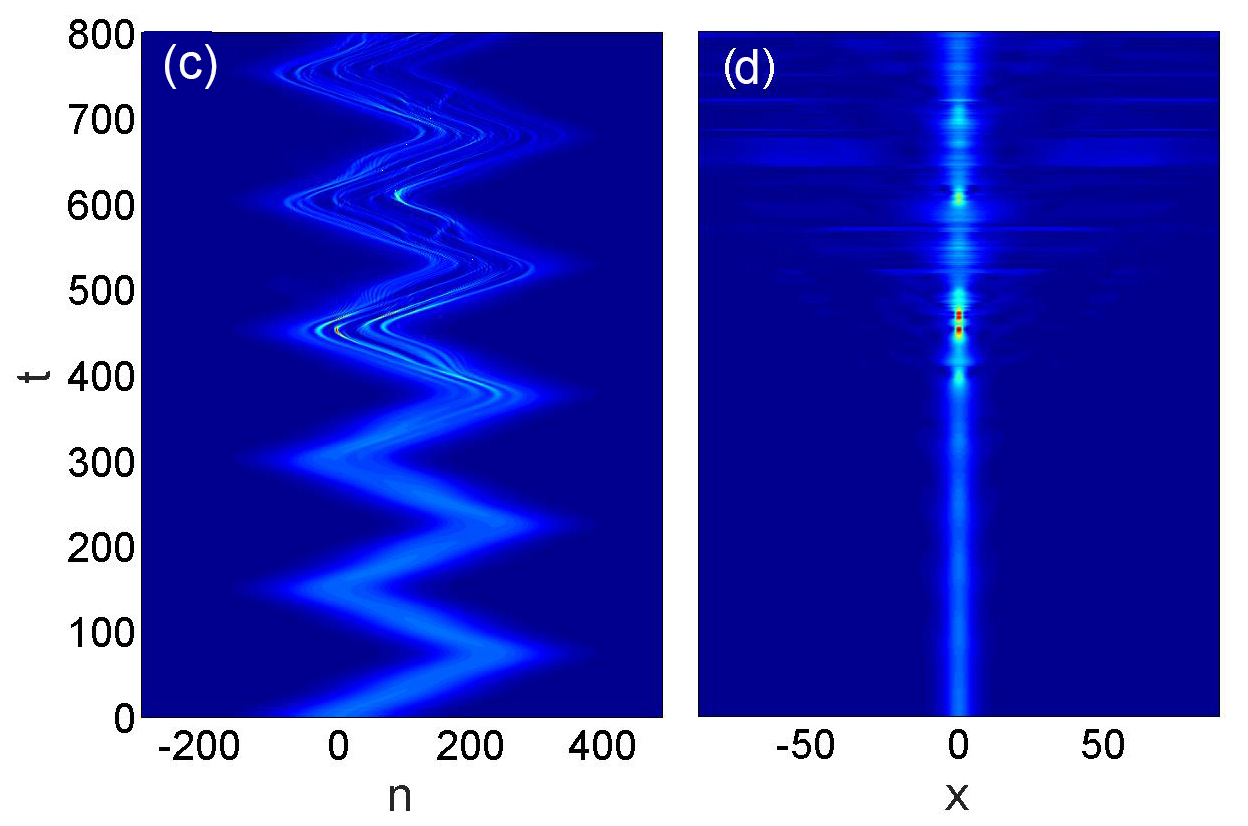}
\caption{(Color online) The long-time evolution of the wave packet in the
nonlinear system in the $\left( n,t\right) $ (a) and $\left( x,t\right) $
(b) planes, respectively. Panels (a) and (b) display the dynamical
localization of the wave packet driven with the frequency corresponding to
the second root of the Bessel function $J_{0}$ [see Eq. (\protect\ref%
{eq:zeroBessel})]. Accordingly, the parameters are chosen as $\protect\gamma %
=0.1$, $\protect\omega _{0}=0.0181$, and $a_{0}=0.15$. Panels (c) and (d)
display the evolution for the same parameters as in (a) and (b), but with a
larger input amplitude, $a_{0}=0.3$.}
\end{figure}

To characterize the dynamics of the wave packet in both the linear and
nonlinear systems, we define the average of a semi-discrete function, $%
f_{n}(x,t)$, carried by the wave packet $u_{n}(x,t)$, as $\langle f\rangle
=P^{-1}\int_{-\infty }^{+\infty }\sum_{n}f_{n}(x,t)|u_{n}(x,t)|^{2}dx$, with
norm $P\equiv \sum_{n}\int_{-\infty }^{+\infty }|u_{n}(x,t)|^{2}dx$. This
definition allows one to explore the average positions of the wave packet
along the and $n$ directions, i.e., $\langle x\rangle $ and $\langle
n\rangle $, respectively. Furthermore, we define a deformation parameter
characterizing \textquotedblleft combined" changes of the wave packet's
widths in the course of the evolution:
\begin{equation}
\Delta (t)=\sqrt{[N(t)-N(0)]^{2}+[X(t)-X(0)]^{2}}\,,  \label{Delta}
\end{equation}%
where the average widths of the wave packet in the $n$ and $x$ directions
are
\begin{equation}
N(t)=\sqrt{\langle (n-\langle n\rangle )^{2}\rangle },~X(t)=\sqrt{\langle
(x-\langle x\rangle )^{2}\rangle }.  \label{NX}
\end{equation}
{If deformations with respect to $n$ and $x$ are strongly anisotropic, $%
\Delta $ estimates the largest one.} For the ideal case of a totally robust
DL, $\Delta (t)$ would remain time independent, while growing {or decreasing}
$\Delta (t)$ corresponds to ongoing deformation of the initial wave packet.
These indicators, pertaining to the dynamical regimes displayed in
Figs.~2(a,b) and 3(a,b), are presented, severally, in three top and three
bottom panels of Fig.~4. It is clearly observed that the growth of the
soliton's width in the course of the long-time evolution is strongly
suppressed. Furthermore, the top panel in Fig. 4 demonstrates that the
soliton may be even slowly shrinking in the $x$ direction. Thus, we conclude that the optimal
nonlinear regime, supporting the robust wave packets in the semi-discrete
(2+1)D model, that was introduced in Ref. \cite{Driben_BlochOscill},
supports the stable DL as well. Also similar characteristics of linear propagation regimes are plotted by dashed curves together with solid curves representing their nonlinear counterparts. We can clearly see that for linear cases the packets spread fast in the X-domain leading to the much faster growth of $\Delta (t)$. For the comparison all the parameters besides the switched off nonlinear interaction were taken similar.

%\begin{figure}[tbp]
%\includegraphics[width=0.5\textwidth]{Table.jpg}
%%  \includegraphics[width=0.5\textwidth]{NonlinearFirstBessel.jpg}
%\caption{$\Delta (t=20000)$, i.e., the spread after a long time evolution,
%as function of the nonlinearity $g$.}
%\end{figure}

Our model also proves its effectiveness in case of the combined ac-dc
modulation of the gradient strength. As mentioned above, in this case the
stabilization occurs at frequencies obeying Eq. (\ref{m}), i.e., they
pertain to zeros of the Bessel functions of the order higher than zero. In
particular, Fig. 5 demonstrates the robust long-time evolution of the wave
packet under the combined ac-dc drive with $\gamma _{\mathrm{dc}}=\omega
_{1}=0.0261$, the other parameters being the same as in Figs. 2(a,b).

\begin{figure}[tbp]
\includegraphics[width=0.5\textwidth]{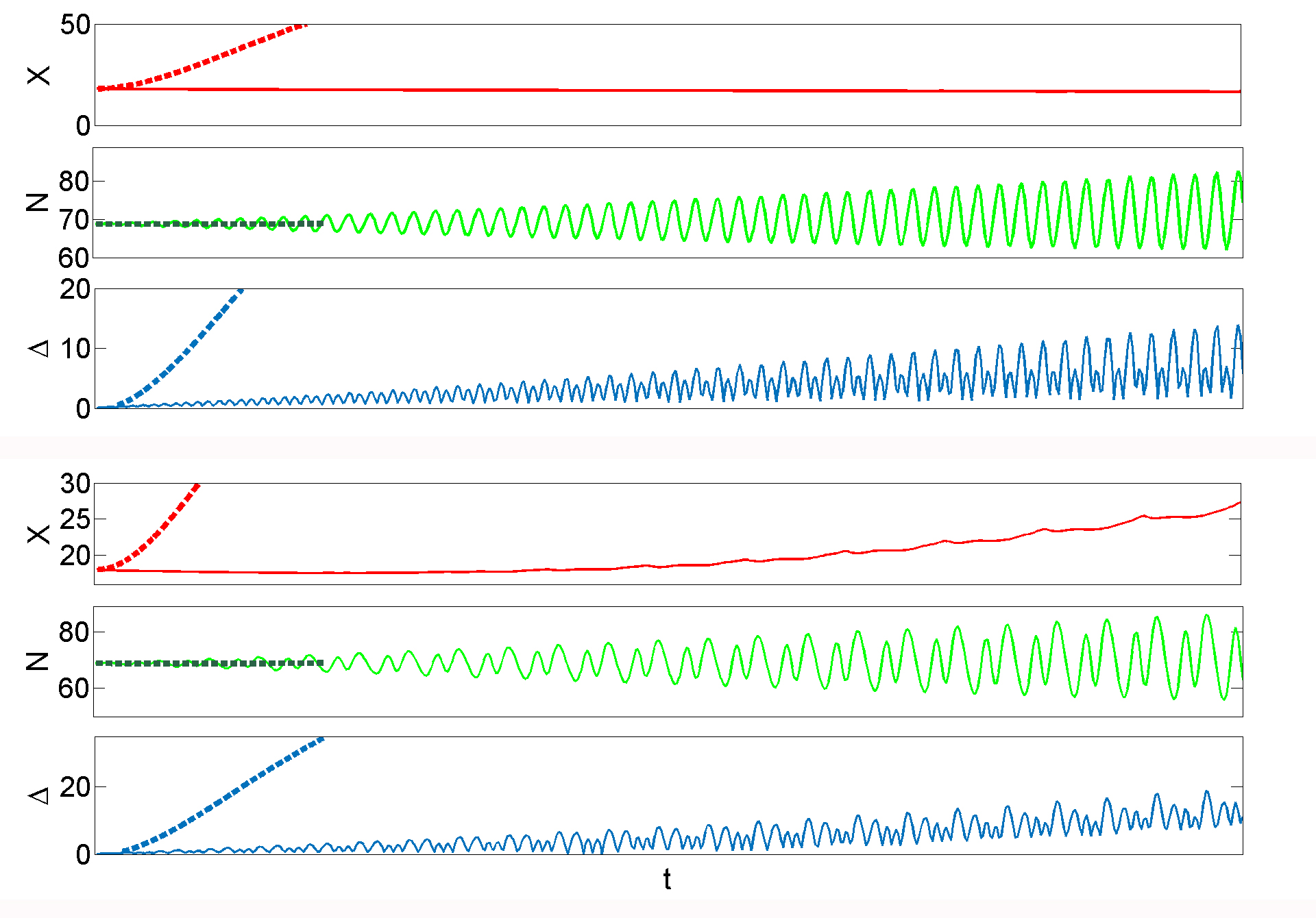}
\caption{(Color online) The temporal evolution of the overall spread of the
wave packet $\Delta (t)$, together with its $N$- and $X$-components [see
Eqs. (\protect\ref{Delta}) and (\protect\ref{NX})], for the case of the DL,
displayed in Figs.~2(a,b) and 3(a,b), are shown here in the three top and
bottom panels, respectively. Dashed curves pertain to characteristics describing linear propagation regimes.}
\end{figure}

\begin{figure}[tbp]
\includegraphics[width=0.5\textwidth]{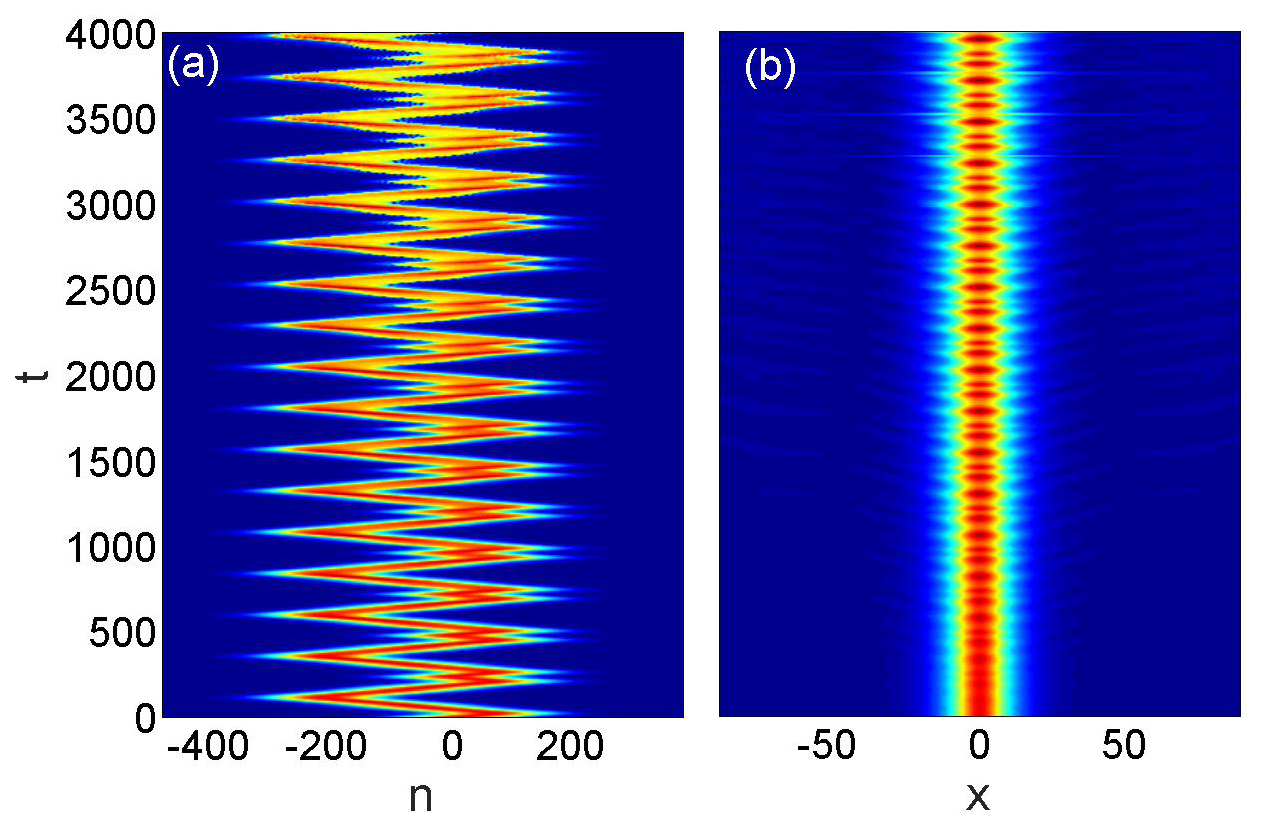}
\caption{(Color online) The evolution of the wave packet in case of the
combined ac-dc modulation of the gradient field, with the frequency
corresponding to the first nontrivial zero of $J_{1}$, see Eq. (\protect\ref%
{m}).}
\label{Fig1}
\end{figure}

\section{Conclusions}

\label{concl}

To study the effect of DL (dynamical localization) in nonlinear settings, we
have introduced a semi-discrete 2D system, driven by the gradient potential,
applied in the discrete direction, which is subject to the time-periodic
modulation. In the absence of the nonlinearity, direct simulations
demonstrate straightforward spreading of localized inputs, even if they
satisfy the specific DL\ condition. The situation is found to be altogether
different in the system with the cubic nonlinearity: in a certain range of
the nonlinearity strength, the system features robust self-trapping of
well-localized wave packets. In the presence of the nonlinearity, similar
DL\ effects are also produced by the application of the gradient potential
subject to the combined dc-ac temporal modulation. The DL effects predicted
by the present analysis may be implemented in an effectively 2D BEC, loaded
into a quasi-1D\ lattice potential, as well as in optics, in the temporal
and spatial domains alike, using, respectively, an array of nonlinear fibers
or a stack of planar waveguides.

%A challenging possibility is to extend the system and its analysis to the 3D
%setting, with one continuum and two transverse discrete directions, which
%may be implemented in BEC, using a deep quasi-2D lattice potential, or using
%available optical bulk waveguides with quasi-2D photonic lattices written in
%them \cite{Jena,Jena2}. In this case, various scenarios for the DL of 3D
%wave packets may be expected.

\begin{acknowledgements}
T.M. and R.D. acknowledge support of the DFG
(Deutsche Forschungsgemeinschaft) through the TRR
142 (project C02) and thank the PC2 (Paderborn Center
for Parallel Computing) for providing computing time.
The work of B.A.M. is partly supported by by the joint
program in physics between NSF and Binational (US-Israel) Science Foundation
through project No. 2015616, and by the Israel Science Foundation through Grant No. 1286/17.
The work of AVY was financially supported by the Government of the Russian Federation
(Grant 074-U01) through ITMO Fellowship scheme.
\end{acknowledgements}

\end{document}